\documentclass[pra,aps,reprint,superscriptaddress]{revtex4-2}

\usepackage{amsmath} 
\usepackage{amssymb}
\usepackage{braket}
\usepackage{bm}
\usepackage[dvips]{epsfig}
\usepackage{graphicx} 

\usepackage{color} 

\newcommand{\comment}[1]{}

\begin{document}

\title{Dissipative search of an unstructured database  } 

\author{Armen E. Allahverdyan}
\affiliation{A. Alikhanyan National Science Laboratory (YerPhI), 0036 Yerevan, Armenia} 

\author{David Petrosyan}
\affiliation{A. Alikhanyan National Science Laboratory (YerPhI), 0036 Yerevan, Armenia} 
\affiliation{Institute of Electronic Structure and Laser, FORTH, GR-70013 Heraklion, Crete, Greece} 

\begin{abstract} The search of an unstructured database amounts to
finding one element having a certain property out of $N$ elements.  
The classical search with an oracle checking one element at a time requires
on average $N/2$ steps.  The Grover algorithm for the quantum search,
and its unitary Hamiltonian evolution analogue, accomplish the search
asymptotically optimally in $\mathcal{O} (\sqrt{N})$ time steps.  
We reformulate the search problem as a dissipative, incoherent Markov process 
acting on an $N$-level system weakly coupled to a thermal bath. Assuming that the
energy levels of the system represent the database elements, we show
that, with a proper choice of the spectrum and long-range but bounded transition rates 
between the energy levels, the system relaxes to the ground state, corresponding to the sought element, 
in time $\mathcal{O} (\ln N)$.  
\end{abstract}

\comment{
Conference abstract:

The search of an unstructured database amounts to finding one element
having a certain property out of $N$ elements. The classical search
with an oracle checking one element at a time requires on average $N/2$
steps.  The Grover algorithm for the quantum search, and its unitary
Hamiltonian evolution analogue, accomplish the search asymptotically
optimally in $O(\sqrt{N})$ time steps. We reformulate the quantum search
as a problem in non-equilibrium statistical mechanics \cite{aa}. Here
the search amounts to relaxing to the ground state for a quantum system
having a specific, gapped energy spectrum. We show that, under proper,
but not too restrictive, conditions, a dissipative Markov process in the
$N$-level system coupled to a thermal bath leads to system's relaxation
to the ground state during time $O(\ln N)$ \cite{aa}. 

Motivation letter for PRL: 
We show that, under proper, but not too restrictive, conditions, 
a dissipative Markov process in an N state system coupled to a 
thermal bath leads to system's relaxation to the ground state 
during time O(ln N). The system is analogous to an unstructured 
set of N elements, with the ground state representing the sought 
element. The classical search time in an unstructured database 
scales as O(N), while the optimal quantum search time scales 
as O(\sqrt{N}). Identifying the relaxation time of our system 
with the search time, we thus obtain exponentially better scaling 
with the system size N. 
}

\date{\today}

\maketitle

\section{Introduction}
In a classical search of an unstructured set of $N$ elements, finding an element 
with a specific feature -- verified by some function (oracle) applied 
to each element at a time -- involves on average $N/2$ steps \cite{comment_unstructured}.
One of the hallmarks of quantum computation is the Grover search algorithm which yields 
quadratic speedup of the search time $\tau = \mathcal{O}(\sqrt{N})$ \cite{grover,review}. 
The quantum search can be formulated as a Hamiltonian evolution of an analog quantum system \cite{farhi}.
The $N$ elements of the set are associated with the orthonormal basis states $\{\ket{w_k} \}_{k=1}^N$,
one state $\ket{w_\ell}$ corresponding to the sought element having the energy $\epsilon \neq 0$,
while the energies of all the other states $\ket{w_{k \neq \ell}}$ being zero. The Hamiltonian
of the system is then
\begin{equation}
H_{0} = \epsilon \ket{w_\ell} \bra{w_\ell}, \label{eq:hamo}
\end{equation}
where $\epsilon$ is a known constant, but $\ell$ is not known. To find $\ell$, 
and thereby $\ket{w_\ell}$, one prepares the system in the equally-weighted superposition
state $\ket{\psi(0)} = \ket{s} \equiv \frac{1}{\sqrt{N}}\sum_{k=1}^N \ket{w_k}$ \cite{comment_sstate} 
and adds to $H_{0}$ the interaction Hamiltonian $V = \epsilon \ket{s}\bra{s}$
that couples all the basis states, 
\begin{equation}
\braket{w_l| V |w_k} = \frac{\epsilon}{N} \quad \forall \, l,k. \label{eq:bobo} 
\end{equation}
The system evolution is governed by the Schr\"odinger equation,
and at time $\tau = \frac{\pi \hbar}{2 |\epsilon|}\sqrt{N}$ the system attains 
the desired state $\ket{\psi(\tau)} = \ket{w_{\ell}}$ \cite{farhi}.
The time $\tau$ does not depend on $\ell$, and the evolution should be terminated 
immediately thereafter in order for the target state $\ket{w_\ell}$ to be correctly identified. 
The asymptotic scaling of the search time $\tau = \mathcal{O}(\sqrt{N})$ 
is optimal for the coherent quantum evolution \cite{review}, 
and it can also be deduced from the time-energy uncertainty relation \cite{tamm} 
applied to the Hamiltonian $H_0+V$ acting on $\ket{\psi(0)}$.

Here we demonstrate an exponential speedup of the search by
using, instead of the coherent Schr\"odinger dynamics that generates
unitary evolution, a dissipative Markov dynamics with long-range but bounded
transition rates between the energy levels. This system differs from the standard 
quantum computation paradigm in terms of computational resources (see Appendix~\ref{app:s12}). 
We begin again with the Hamiltonian (\ref{eq:hamo}), assuming that $\epsilon < 0$ and 
therefore $\ket{w_\ell}$ is the ground state of $H_{0}$.
We then add an auxiliary (known) Hamiltonian $H_1$ which lifts the degeneracy 
of all the energy levels $\ket{w_k}$, but still leaves $\ket{w_\ell}$
as a ground state of $H = H_0 + H_1$ for any $\ell$, with some energy gap $\Delta_\ell$.
We next couple the system to a thermal bath at a temperature $T=1/\beta$ ($k_{\rm B}=\hbar=1$)
and let it relax to the Gibbsian (equilibrium) state described by the density operator 
$\rho \propto e^{-\beta H}$. If $T$ is sufficiently low, 
such that $\Delta_\ell \beta > 1$ for any $\ell =1,2, \dots, N$, 
then $\rho \approx \ket{w_\ell}\bra{w_\ell}$, and this approximation can be made arbitrary precise 
by increasing the energy gap $\Delta_\ell$ or decreasing the temperature $T$. 

The working time of our dissipative analog device $\tau_{\mathrm{rlx}}$ is proportional to 
the inverse relaxation rate $\alpha^{-1}$ of the system towards the equilibrium state $\rho$, 
and our main goal is to estimate this time, given the spectrum of the system and 
its coupling to the thermal reservoir, as detailed below. 
In contrast to the coherent quantum search, however, we need not demand that $\tau_{\mathrm{rlx}}$ 
be independent on the unknown index $\ell$, because once the system reaches the equilibrium state 
$\rho \approx \ket{w_\ell}\bra{w_\ell}$, it will remain in that state thereafter. 
We can then take the search time as $\tau_{\mathrm{rlx}} = \max_{\ell}[1/\alpha(\ell)]$. 

\section{The system}

We now turn to a more quantitative description of the system. 
The Hamiltonian $H_1$ that shifts the energy levels $\ket{w_k}$
should leave $\ket{w_\ell}$ as a unique ground state of $H = H_0 + H_1$ for any $\ell$. 
Assuming for simplicity that $H_1$ commutes with $H_0$, we therefore require that 
$\mathrm{spectrum}[H_1] = \{\eta_k\}_{k=1}^N$ satisfies the condition
\begin{equation}
\eta_N-\eta_1 < |\epsilon| , \label{eq:H1spcetcond}
\end{equation}
where $\eta_N = \max \{\eta_k\}$ and $\eta_1 = \min \{\eta_k\}$. 
We thus have $\mathrm{spectrum}[H] = \{ \varepsilon_k \}_{k=1}^N = \{\epsilon + \eta_\ell, \, \{\eta_k\}_{k\neq \ell}\}$
with the smallest energy gap between state $\ket{w_\ell}$ and all the other states $\ket{w_k}$ being 
$\Delta_N =  \eta_1 - (\eta_N + \epsilon) > 0$.

The (incoherent) Markov dynamics of the system weakly coupled to a thermal bath is governed by the rate
equations for the populations $p_k$ of states $\ket{w_k}$ (see Appendix \ref{app:s1}):
\begin{equation}
\dot{p}_k = \sum_{l=1}^N v_{kl} p_l - p_k \sum_{l=1}^N v_{lk} , \label{eq:markov}
\end{equation}
where $v_{kl} \geq 0$ ($v_{kk} = 0$) are the transition rates between the energy levels 
induced by the thermal bath at temperature $T$. Hence, we have the detailed balance condition
\begin{equation}
v_{kl}e^{-\beta \varepsilon_l}=v_{lk}e^{-\beta \varepsilon_k}, \label{eq:detbal}
\end{equation}
which implies that there is a unique stationary state of the system determined by 
the Gibbs (equilibrium) probabilities $p_k = Z^{-1} e^{-\beta\varepsilon_k}$ for $k=1,2,\ldots , N$,
with $Z = \sum_{k=1}^N e^{-\beta\varepsilon_k}$. 
In view of the spectrum of the Hamiltonian $H$, the probability of the ground state is
\begin{equation}
p^{(\mathrm{eq})}_1 =
\left[ 1 - e^{\beta\epsilon} \left(e^{\beta\eta_\ell} {\sum}_{k=1}^N e^{-\beta\eta_k} - 1 \right)\right]^{-1},
\label{eq:mont}
\end{equation} 
and $P_{\mathrm{err}} = 1-p^{(\mathrm{eq})}_1$ is then the error probability of our dissipative search. 
Obviously, the smallest ground state probability in the stationary state, and thus the largest error,
would occur for $\ell = N$ with the ground state having the largest possible energy $\epsilon + \eta_N$, 
see Eq.~(\ref{eq:H1spcetcond}). To ensure the ground state dominance, we therefore require that
$e^{\beta|\epsilon|} \gg \big(e^{\beta\eta_N} \sum_{k=1}^N e^{-\beta\eta_k} - 1 \big)$.

We are interested in the dependence of the relaxation time to the equilibrium state on $N$.  
We therefore demand that the total transition rate from any state $\ket{w_k}$ be bounded,
\begin{equation}
\gamma_k \equiv \sum_{l=1}^{N} v_{lk} \lesssim v \quad \forall \; k, 
\label{eq:cri}
\end{equation}
where $v$ is some constant independent on $N$ [see Eq.~(\ref{eq:Grauberv}) below]. 
This condition determines the physically acceptable transition rates $v_{kl}$ and  
excludes parallel relaxation processes that would lead to a trivial acceleration of the dynamics. 
\comment{Note that similar condition also holds for the Hamiltonian quantum search, where all the states
are coupled to each other with $V_{kl} = \epsilon/N$, Eq.~(\ref{eq:bobo}).}

Condition (\ref{eq:cri}) automatically holds for short-range transitions: 
$v_{kl} \sim v \neq 0$ only for $|k-l| \leq r$ with some fixed range $r$ that does not depend on $N$. 
But then, starting from any arbitrary state $\ket{w_k}$, we will reach the desired ground state
$\ket{w_\ell}$, where the population accumulates, via diffusive
transport, and the relaxation time will scale as $\tau_{\mathrm{rlx}} =
\mathcal{O}(N^2)$ \cite{kampen,brandes}.  This is worse than using either
ballistic transport or classical search with the $\tau = {\cal O}(N)$
scaling.  We shall therefore consider long-range but bounded (\ref{eq:cri}) 
transition rates $v_{kl} > 0 \, \forall \, k \neq l$. 

We can rewrite Eq.~(\ref{eq:markov}) for the vector $p = [p_1,p_2,\ldots,p_N]^{\mathrm{T}}$ in a matrix form, 
\begin{equation}
\dot{p} = A \, p, \quad A_{kl} \equiv v_{kl} - \delta_{kl} \gamma_k , \label{eq:ME}
\end{equation}
where $A_{k\not=l}\geq 0$ and $\sum_{k=1}^N A_{kl}=0$.
If follows from Eq.~(\ref{eq:detbal}) that $\tilde{A}_{kl}=A_{kl}e^{-\frac{\beta}{2}(\epsilon_l-\epsilon_k)}=\tilde{A}_{lk}$
is a symmetric matrix that has the same eigenvalues as $A_{kl}$. 
Hence the eigenvalues of $A$ are real. 
Assuming that $A$ is imprimitive, $[A^{m}]_{k\not=l}>0$ for a sufficiently large integer $m$, 
we can employ the Perron-Frobenius theorem \cite{kampen,schna,perron} to argue that 
the largest eigenvalue $\alpha_1$ of $A$ is zero and non-degenerate.
Hence, all the other eigenvalues are negative, $\alpha_1 =0 > \alpha_2 \geq \alpha_3, \ldots$
Note that $0 \geq \alpha_2 \geq \alpha_3, \ldots$ follows directly
from the much simpler Gershgorin's circle theorem \cite{schna}. 
The largest non-zero eigenvalue $\alpha_2$ of $A$ defines the (exponential) relaxation time 
$\tau_{\mathrm{rlx}} = 1/|\alpha_2|$ of the master equation (\ref{eq:ME}) (see Appendix \ref{app:s2}). 

The detailed balance condition (\ref{eq:detbal}) and the requirement that $A$ be imprimitive still leaves 
some freedom in choosing the transition rates $v_{kl}$. A well-known phenomenological approach is 
to use the Glauber rates \cite{glauber} widely employed in statistical mechanics \cite{heims,martin}.
We thus assume  
\begin{equation}
v_{kl} = \frac{v}{\max (n_k, n_l)} \, \left(1+e^{-\beta \varepsilon_l+\beta \varepsilon_k} \right)^{-1}, \label{eq:Grauberv}
\end{equation}
where $v$ is the bare relaxation rate that depends on the strength of the system-bath coupling, 
but does not depend on $N$ and on the number of levels $n_k$ with energies not larger than $\varepsilon_k$. 
The physical meaning of Eq.~(\ref{eq:Grauberv}) is that transitions from higher to lower energy levels are facilitated, 
while the reverse transitions are suppressed, with the condition (\ref{eq:detbal}) obviously satisfied. 
The Glauber rates are usually written without the factor $1/\max(n_k,n_l)$, but since we allow transitions 
between all energy levels, this factor is needed to satisfy condition (\ref{eq:cri}). 
The Glauber rates can be deduced from the Born-Markov treatment of weak system--bath coupling 
with an appropriate bath spectrum (see Appendix \ref{app:s11}). 

Consider first the trivial case of $H_1=0$, i.e. all states have the same energy equal to zero,
and the ground state energy is $\epsilon<0$. From Eqs.~(\ref{eq:markov}) and (\ref{eq:Grauberv}), 
we obtain for the ground state population 
\begin{gather*}
\dot{p}_1 = \frac{v}{N (1 + e^{\beta\epsilon})} - \frac{p_1}{\tau_{\mathrm{rlx}}}, \qquad 
\frac{1}{\tau_{\mathrm{rlx}}} = \frac{v}{N}\, \frac{1 + (N-1) e^{\beta\epsilon}}{ 1 + e^{\beta\epsilon}}, 
\end{gather*}
which leads to the equilibrium population $p_1^{(\mathrm{eq})} = [1+(N-1)e^{\beta\epsilon}]^{-1}$
attained exponentially for times $t \gg \tau_{\mathrm{rlx}}$.
Thus the ground state dominance, $p_1^{(\mathrm{eq})} \simeq 1$ for $(N-1)e^{\beta\epsilon}\ll 1$, 
leads to $\tau_{\mathrm{rlx}} = \mathcal{O}(N)$, which is the expected result for the classical search of an unstructured set.  

To obtain more interesting results, consider the auxiliary Hamiltonian $H_1$ with a non-degenerate spectrum
\begin{equation}
\eta_k = a \ln(k), \quad k=1,...,N,  \label{eq:H1spect}
\end{equation}
where $a > 0$. Hence, the energies $\eta_k$ grow only slowly (logarithmically) with index $k$, 
which justifies the possibility of long-range coupling between the energy levels. 
We then set $\epsilon = - b\ln(N)$ with the parameter $b>a$, to satisfy the condition (\ref{eq:H1spcetcond}).  
In Appendix \ref{app:s3} we present explicit expressions for the Glauber rates for the logarithmic 
spectrum (\ref{eq:H1spect}) and show that condition (\ref{eq:cri}) also holds.
The ground state dominance, $p_1^{(\mathrm{eq})} \simeq 1$, now requires that 
$N^{b\beta} \gg \big( N^{a\beta} \sum_{k=1}^{N} k^{-a\beta} - 1 \big)$, which for $N\gg 1$ is satisfied 
if $b\beta> \max [1,\,a\beta]$, see Eq.~(\ref{eq:mont}). 
Indeed, for $a\beta>1$ we have $N^{a\beta}\sum_{k=1}^{N}k^{-a\beta}= \mathcal{O}(N^{a\beta})$, 
while for $a\beta<1$ we have $N^{a\beta}\sum_{k=1}^{N}k^{-a\beta}=\mathcal{O}(N)$.   
The error probability 
\begin{equation}
P_{\mathrm{err}} = 1-p_1^{(\mathrm{eq})} \simeq N^{-(b-a)\beta} \sum_{k=1}^{N} k^{-a\beta}
\end{equation}
can be arbitrary small, $P_{\mathrm{err}}  \approx  N^{-(b-a)\beta}$, when $a \beta >1$.

\begin{figure}[t]
\includegraphics[width=1.0\columnwidth]{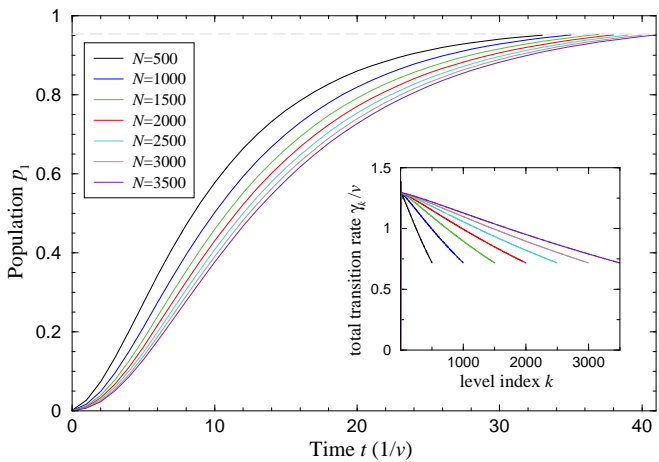}
\caption{Dynamics of population of the ground state $p_1$ (time $t$ is in units of $v^{-1}$)
for various $N = 500,1000,\ldots,3500$ (progressively lower curves), 
as obtained from the numerical solution of Eqs.~(\ref{eq:markov}) for the system with 
logarithmic spectrum (\ref{eq:H1spect}) and the corresponding Glauber rates (\ref{eq:Grauberv}),
with the parameters $a \beta =1.2$ and $b \beta = 2$. 
The initial populations are $p_k =1/N \; \forall \; k \in [1,N]$, the sought index is $\ell = N$, 
i.e., the ground state energy is $\varepsilon_1 = (a-b) \ln N$, and the dynamics is terminated
once $p_1 \geq 0.95$. Inset shows the transition rates $\gamma_k$ of Eq.~(\ref{eq:cri}).
For each curve, the index $k$ runs from 1 to the corresponding $N$. Curves from top to bottom on the
main figure have the same value of $N$ as curves from left to right in the inset, e.g. the top curve on the
main figure and the leftmost curve on the inset refer to $N=500$.   }
\label{fig:dynamics}
\end{figure} 

In Fig.~\ref{fig:dynamics} we show the dynamics of population of the ground state $p_1$,
as obtained from numerical solutions of the rate equations (\ref{eq:markov}), with the
initial population equally distributed among all the energy levels, while the sought index
is $\ell = N$ corresponding to the smallest energy gap $\Delta_N = (b-a) \ln N$. 
We observe that the time at which the ground state population exceeds some threshold value, 
e.g. $p_1 \geq 0.95$, grows very slowly with increasing the system size $N$.
In the inset of Fig.~\ref{fig:dynamics} we show the total transition rates $\gamma_k$
from levels $k$, which remain bounded for any $N$ as per condition (\ref{eq:cri}).

\begin{figure}[t]
\includegraphics[width=1.\columnwidth]{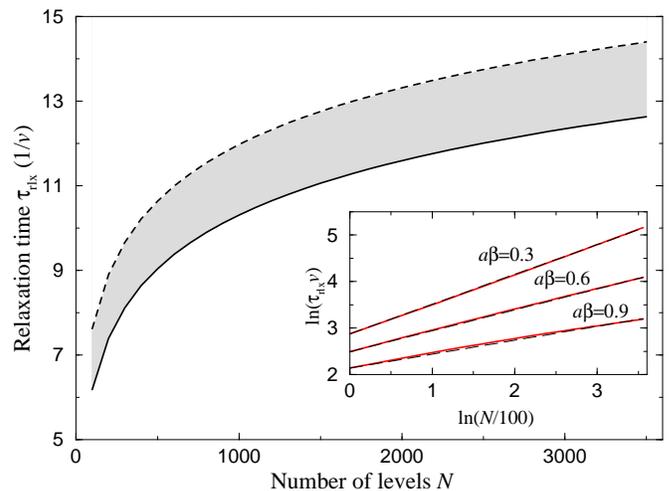}
\caption{Relaxation time $\tau_{\mathrm{rlx}} = 1/|\alpha_2|$ (in units of $v^{-1}$) 
versus $N$, as obtained from diagonalization of matrix $A$ in Eq.~(\ref{eq:ME}) 
for the system with logarithmic spectrum (\ref{eq:H1spect}) and the corresponding 
Glauber rates (\ref{eq:Grauberv}), with $a \beta =1.2$ and $b \beta = 2$. 
For $\ell = N \gg 1$ (solid line) the relaxation time is well approximated by
$\tau_{\mathrm{rlx}} v \approx 1.82 \ln N - 2.3$, and  
for $\ell = 1$ (dashed line) by $\tau_{\mathrm{rlx}} v \approx 1.92 \ln N -1.34$.
Nearly indistinguishable curves are obtained for $(a\beta, b\beta)=(3, 4), (1.2, 4), (1.2, 5)$. 
Inset shows $\ln (\tau_{\mathrm{rlx}} v)$ versus $\ln(N/100)$ for smaller values of 
$a\beta$ (red solid lines) and $b\beta =2$, with each line approximated by
$\mathrm{const} + \kappa \ln(N/100)$ (black dashed lines) corresponding to a power-law behavior 
$\tau_{\mathrm{rlx}} \propto N^{\kappa}$ with $\kappa=0.64,0.45,0.3$ for $a\beta=0.3, 0.6, 0.9$,
respectively.}
\label{fig:taurlx}
\end{figure}

In Fig.~\ref{fig:taurlx} we show the relaxation time $\tau_{\mathrm{rlx}} = 1/|\alpha_2|$ as
a function of $N$, obtained from the diagonalization of matrix $A$ in Eq.~(\ref{eq:ME}).
For $a\beta > 1$, $b>a$ and large $N \gg 1$, the relaxation time growth logarithmically with $N$ 
\begin{equation}
\tau_{\mathrm{rlx}} = \mathcal{O} (\ln N) . \label{eq:tauln}
\end{equation}
Note that $\tau_{\mathrm{rlx}}$ depends only weakly on the sought index $\ell$, which determines the energy gap $\Delta_{\ell}$
but also the spectrum of the excited states and its bandwidth, especially for small $\ell$.  
But once the ground state dominance condition is satisfied, the relaxation time does not change 
upon increasing $(b-a)$.

The relaxation time $\tau_{\mathrm{rlx}}$ increases for smaller values of $a\beta$, since then the transition
rates between excited levels tend to equalize, e.g. for $\ell=N$ we have
$v_{kl}/v_{lk}=(l/k)^{a\beta}$ for $l>k>1$ (see Appendix~\ref{app:s3}).
In other words, for smaller $a\beta$, the system wanders longer among
the excited levels before relaxing to the ground state. 

For $a\beta$ sufficiently smaller than $1$, the relaxation time follows a power-law \cite{hilhorst}
\begin{equation}
\tau_{\mathrm{rlx}} = \mathcal{O} (N^\kappa) , \label{eq:taupl}
\end{equation}
with the exponent $\kappa < 1$ that depends on $a$, see the inset of Fig.~\ref{fig:taurlx}. 
For $a \to 0$, we approach $\kappa \to 1$ of the classical search time, 
while the error probability is $P_{\mathrm{err}} \sim N^{-(b-1)\beta}$.
The transition from the logarithmic (\ref{eq:tauln}) to the power-low (\ref{eq:taupl}) dependence
of $\tau_{\mathrm{rlx}}$ on $N$ is gradual taking place in the vicinity of $a\beta \simeq 1$.  
We emphasize that $\kappa<0.5$, as shown in the inset of Fig.~\ref{fig:taurlx}, 
already signifies better scaling of the dissipative search time with $N$ than that 
of the unitary (Grover) search. 

\section{Conclusions}

To summarize, we have shown that a dissipative Markov dynamics in a system with 
a weakly non-degenerate spectrum of $N \gg 1$ states can result in the relaxation 
of the system to the (unknown) ground state during time $\tau_{\mathrm{rlx}} = \mathcal{O} (\ln N)$. 
The system can be viewed as an analog of an unstructured database of $N$ elements, 
for which the classical search time scales as $\mathcal{O}(N)$ while the 
optimal quantum search time scales as $\mathcal{O}(\sqrt{N})$. 
We can identify the relaxation time of our system with the search time that has exponentially
better scaling with the system size $N$ than either the classical or the fully quantum search. 

The necessary condition for achieving the short relaxation times of the
system, apart from the (weakly) non-degenerate spectrum, is that the
Markov process involves transitions between arbitrary energy levels.
We should ensure, however, that this long-range interactions are bounded for
any $N$, since otherwise decreasing the search time with increasing $N$ would be
trivial.  Note that long-range coupling between energy levels (\ref{eq:bobo})
is also present in the Hamiltonian analog of quantum search \cite{farhi}. 

Since the Markov dynamics of Eqs.~(\ref{eq:markov}) is described in terms 
of classical probabilities, it is natural to ask whether the considered
dissipative search can be implemented on a classical computer using,
e.g. Monte-Carlo simulations to reach the equilibrium state dominated by
the ground state. It can of course be done, but will require large
amount of calculations and computer memory.  Indeed, the $N=2^n$ energy
levels may correspond to different configurations of $n$ bits or spins
$\{\sigma_i=\pm 1\}_{i=1}^n$.  Recall that the energy levels in our
dissipative analog device should be (weakly) non-degenerate, which means
that we need to realize a weakly-interacting $n$-spin system. We will
then have to calculate $\sim N$ different energies for $2^n$ different
configurations and store them in the memory, in order to determine the
transition probabilities between the different energy levels. And these
transitions may involve up to $n$ simultaneous spin-flips
($\sigma_i\to-\sigma_i$).  In contrast, in the usual Monte-Carlo
simulations only one spin is flipped at a time, and the energy
difference between the old and new configurations is easy to calculate
at each time step. 

In our study, we assumed weak system-bath coupling and employed the
Glauber rates \cite{glauber,heims,martin} for the transitions between
the energy levels. But our results equally hold for other similar
coupling schemes, e.g. Arrhenius rates often employed in chemical
physics \cite{kampen,hilhorst}. We note that master equations with
long-range transition rates are frequently employed for describing
glassy systems and amorphous materials \cite{hilhorst,trap_review}.
Such models often produce results that agree with the experiments, 
but the relaxation times scale polynomially with $N$. 
A proof of principle demonstration of dissipative search can be realized with 
multilevel molecular systems with incoherently coupled subset of ro-vibrational levels \cite{AtMolSpct}, 
or with atomic Rydberg systems \cite{RydbergAts} with a properly tailored broadband 
(microwave) field that induces transitions between a number of Rydberg energy levels,
with rates that mimic those in Eq.~(\ref{eq:Grauberv}). 
Finally, our results may have important and interesting implications 
for protein folding and similar problems, where macromolecules attain 
the target (minimal energy) conformations very fast, despite the available 
huge energy landscape \cite{zwanzig,ah}.

\begin{acknowledgments}
We thank D. Karakhanyan, N.H. Martirosyan and Sevag Gharibian for useful discussions.
This work was supported by SCS of Armenia, grant No. 20TTAT-QTa003. 
A.E.A. was partially supported by a research grant from the Yervant Terzian 
Armenian National Science and Education Fund (ANSEF) based in New York, USA. 
D.P. was also supported by the EU QuanERA Project PACE-IN (GSRT Grant No. T11EPA4-00015).
\end{acknowledgments}

\appendix

\section{Derivation of the Markovian master equation for a system weakly coupled to a thermal bath.}
\label{app:s1}

We consider a system with Hamiltonian $H_{\mathrm{S}}$, a thermal bath with Hamiltonian $H_{\mathrm{B}}$, and their interaction described by Hamiltonian $V_{\mathrm{I}}$.
We assume that the initial state of the system $\rho_{\mathrm{S}}(0)$  is diagonal in the energy representation and we are interested only in the dynamics of populations of its
energy eigenstates $\{\ket{\varepsilon_k}\}$. We further assume that the bath is initially in the thermal equilibrium (Gibbsian) state $\rho_{\mathrm{B}}(0)=\frac{1}{Z}\,e^{-\beta H_B}$, 
$Z = \mathrm{tr} [e^{-\beta H_B}]$, at inverse temperature $\beta = 1/T$, and we set for brevity $k_{\rm B}=\hbar=1$. The system and the bath are initially uncorrelated and their
combined state is 
\begin{equation}
\label{log1}
\rho_{\mathrm{SB}}(0)=\rho_{\mathrm{S}}(0)\otimes \rho_{\mathrm{B}}(0), 
\end{equation}
We can represent the time evolution generated by the Hamiltonian of the composite system $H=H_{\mathrm{S}}+H_{\mathrm{B}}+V_{\mathrm{I}}$ as
\begin{subequations}
\begin{eqnarray}
&& e^{-itH} = \mathcal{T}^* e^{-i\int_0^t d t' \, \tilde{V}_{\mathrm{I}}(-t')}\, e^{-it(H_{\mathrm{S}} + H_{\mathrm{B}})}, \\
&& \tilde{V}_{\mathrm{I}}(-t') \equiv e^{-it' (H_{\mathrm{S}}+H_{\mathrm{B}})} V_{\mathrm{I}} \, e^{it' (H_{\mathrm{S}}+H_{\mathrm{B}})},\\
&& U_t \equiv \mathcal{T}^* e^{-i \int_0^t d t' \,  \tilde{V}_{\mathrm{I}}(-t')},
\label{k7}
\end{eqnarray}
\end{subequations}
where $\mathcal{T}^*$ denotes the time anti-ordering.

Given the eigenresolution of the system Hamiltonian 
$H_{\mathrm{S}}=\sum_{k=1}^N \varepsilon_k|\varepsilon_k\rangle\langle\varepsilon_k|$ with $\langle \varepsilon_k|\varepsilon_l\rangle=\delta_{kl}$,
and using the assumed commutative initial states
\begin{equation}
[H_{\mathrm{S}},\rho_{\mathrm{S}}]=0, \qquad [H_{\mathrm{B}},\rho_{\mathrm{B}}]=0,
\label{bora}
\end{equation}
we deduce from $\rho_{\mathrm{SB}}(t)=e^{-itH}\rho_{\mathrm{SB}}(0) \, e^{itH}$ that
\begin{equation}
\label{me}
p_k(t)\equiv\langle \varepsilon_k |\rho_{\mathrm{S}}(t) |\varepsilon_k\rangle =\sum_{l=1}^N Q_{kl}(t)p_l(0), 
\end{equation}
where 
\begin{equation}
\label{mesh}
Q_{kl}(t) = \langle \varepsilon_k |{\rm tr}_{\mathrm{B}} [U_t |\varepsilon_l\rangle
\langle \varepsilon_l |\otimes \rho_{\mathrm{B}}(0) U_t^\dagger ] |\varepsilon_k\rangle .
\end{equation}
is a stochastic, i.e. probability, matrix: $Q_{kl}(t)\geq 0$ and $\sum_{k=1}^N Q_{kl}(t)=1$.
Hence, Eqs.~(\ref{me}) and (\ref{mesh}) describes a classical (though generally non-Markovian) process.  

Note that during the evolution the non-diagonal components (coherences) of the system density matrix
$\langle \varepsilon_k |\rho_{\mathrm{S}}(t) |\varepsilon_l\rangle$ may in general be non-zero, 
even though they were assumed $\langle \varepsilon_k |\rho_{\mathrm{S}}(0) |\varepsilon_l\rangle =0$ ($k \neq l$) for $t=0$. 
Yet, the coherences $\langle \varepsilon_k |\rho_{S}(t) |\varepsilon_l\rangle$ do not explicitly enter Eq.~(\ref{me}) 
[their contribution is contained in $Q_{kl}(t)$]. 
We therefore need not invoke the rotating-wave approximation \cite{lindblad,breuer,farina}, 
which would be inapplicable for a system having many densely spaced energy levels. 

To make the Markov approximation for the transition probabilities $Q_{kl}(t)$, we first expand $U_t$ in Eq.~(\ref{k7}) to second order in $\tilde{V}_I$:
\begin{equation}
U_t = 1-i\int_0^t \!\! d t' \, \tilde{V}_I(-t') - \int_0^t \!\! d t_1 \int_0^{t_1} \!\! d t_2 \, \tilde{V}_I(-t_2) \tilde{V}_I(-t_1) .
\end{equation}
We now consider the interaction Hamiltonian 
\begin{equation}
\label{hamo3}
V_I=g \, S \otimes B, 
\end{equation}
with a weak coupling $g$ between the system $S$ and bath $B$ operators, and assume $\mathrm{tr}_{\rm B} [B \, \rho_{\mathrm{B}}(0)]=0$.
We then obtain 
\begin{widetext}
\begin{gather}
\label{gret}
Q_{kl}(t) = \delta_{kl} + q_{kl}(t) \\
\label{gre}
q_{kl}(t) \equiv -g^2\delta_{kl} \int_0^t \!\! d t_1 \int_0^{t_1} \!\! d t _2  \left[ \langle \varepsilon_k | S(-t_2) S(-t_1)| \varepsilon_k \rangle K(t_1-t_2) 
+ \langle \varepsilon_k | S(-t_1) S(-t_2)| \varepsilon_k \rangle K(t_2-t_1) \right] \nonumber\\
 + g^2\int_0^t \int_0^{t} \!\! d t_1 d t_2 \,\langle \varepsilon_k| S(-t_1) |\varepsilon_l\rangle\langle \varepsilon_l| S(-t_2)| \varepsilon_k \rangle K(t_1-t_2),
\label{greta}
\end{gather}
where for the bath correlation function $K(t)$ we used its stationary property $K(t-t')=\mathrm{tr}_{\rm B}[B(t) B(t') \, \rho_{\rm B}(0)] = K^*(t'-t)$.
We can collect back the matrix exponent writing Eq.~(\ref{gret}) as
\begin{equation}
Q(t)=e^{q(t)} .
\label{vaza}
\end{equation}
We then calculate 
\begin{gather}
\dot{q}_{kl}(t)= -\delta_{kl} g^2 \sum_{m=1}^N \left[ \langle \varepsilon_k | S | \varepsilon_m \rangle \langle \varepsilon_m | S |\varepsilon_k \rangle \kappa_{km}^{(+)} 
+ \langle \varepsilon_k | S |\varepsilon_m \rangle \langle \varepsilon_m | S |\varepsilon_k \rangle \kappa_{mk}^{(-)} \right] \nonumber \\
+ g^2 \left[ \langle \varepsilon_k| S |\varepsilon_l \rangle \langle \varepsilon_l | S | \varepsilon_k \rangle \kappa_{kl}^{(-)} 
+\langle \varepsilon_k| S |\varepsilon_l \rangle  \langle \varepsilon_l | S |\varepsilon_k \rangle \kappa_{lk}^{(+)} \right] , \\ 
\kappa_{kl}^{(\pm)} =\int_0^t d t' \, e^{it' (\varepsilon_k - \varepsilon_l )} K(\pm t').
\label{k2}
\end{gather}
\end{widetext}
We next assume that the bath correlations $K(t)$ decay with time $t$ sufficiently quickly.
Indeed, the characteristic time $\hbar/(k_{\rm B}T)$ for $K(t)$ can be much smaller than 
the time $\tau_{\mathrm{rlx}}$ over which the populations $p_k$ change significantly (see below). 
This is due to that $\tau_{\mathrm{rlx}}$ contains a large prefactor $g^{-2}$ and, in general, increases with the number of states $N$. 
Hence, we can make the Markov approximation by extending in Eq.~(\ref{k2}) the limit of integration $t \to \infty$,  
\begin{equation}
\label{k3}
\kappa_{kl}^{(\pm)} \approx \int_0^\infty d t' \, e^{it'(\varepsilon_k - \varepsilon_l)} K(\pm t').
\end{equation}
This means that in Eq.~(\ref{vaza}) $q(t)$ is approximately a linear function of time $t$.
Eventually, in the considered Markov limit Eqs.~(\ref{me}) and (\ref{mesh}) are reduced to the master equation
\begin{gather}
\label{master}
\dot{p}_k = \sum_{l=1}^N v_{kl} p_l - p_k \sum_{l=1}^N v_{lk}, \\
\label{master2}
v_{kl}= g^2 |\langle \varepsilon_k | S |\varepsilon_l \rangle|^2 \kappa_{kl}, \\
\kappa_{kl} = \int d t' \, K(t') e^{-it' (\varepsilon_k - \varepsilon_l)}.
\label{master3}
\end{gather}
Hence, the transition rates $v_{kl}$ are proportional to a symmetric (over $k$ and $l$) factor 
$|\langle \varepsilon_k| S |\varepsilon_l \rangle|^2$ multiplied by $\kappa_{kl}$ that depends 
only on the energy difference $\varepsilon_k - \varepsilon_l$. 
In the main text, we emply, at a phenomenological level, the master equation (\ref{eq:markov}) of the same form. 

Introducing the eigenresolution for the bath Hamiltonian, $H_{\mathrm{B}} = \sum_{i} E_i \mathcal{P}_i$, 
where $\mathcal{P}_i$ are eigenprojectors, $\mathcal{P}_i \mathcal{P}_j = \delta_{ij}$ and $\sum_i \mathcal{P}_i = \hat{1}$,
we find from Eq.~(\ref{master3}) that
\begin{equation}
\kappa_{kl}=
\frac{1}{Z} \sum_{ij} \mathrm{tr}_{\mathrm{B}} ( \mathcal{P}_i B \mathcal{P}_j B \mathcal{P}_i) \, e^{-\beta E_i} \delta[E_i-E_j-(\varepsilon_k-\varepsilon_l)],
\label{grr}
\end{equation}
which clearly shows that $\kappa_{kl} \geq 0$ and satisfy the detailed balance condition of Eq.~(\ref{eq:detbal}):
\begin{equation}
\label{aa2}
\kappa_{kl}=\kappa_{lk}e^{ \beta(\varepsilon_l-\varepsilon_k)}.
\end{equation}
We note that Eqs. (\ref{grr}) and (\ref{aa2}) can also be derived from the Bochners theorem and Kubo-Martin-Schwinger 
conditions \cite{breuer}.

\subsection{Glauber rates}
\label{app:s11}

In the main text, we use the Glauber transition rates [see Eq.~(\ref{eq:Grauberv})] 
\begin{equation}
v_{kl} \propto \left( 1 + e^{-\beta \varepsilon_l + \beta\varepsilon_k}  \right)^{-1}.
\label{glauberr}
\end{equation}
Assuming that in Eq.~(\ref{master2}) $|\langle \varepsilon_k | S |\varepsilon_l \rangle|^2$ only weakly depend on the level indexes $k,l$,
we can use Eq.~(\ref{master3}) to determine the necessary form of the bath correlator to obtain the Glauber rates,
\begin{eqnarray}
K(t)&=&\int\frac{d\omega}{2\pi}\frac{e^{-i\omega t}}{1+e^{-\omega/T}} \nonumber \\
&=& \int\frac{d\omega}{2\pi}\frac{e^{-i \omega (t-i\zeta)}}{1+e^{-\omega/T}} = \frac{T}{2i}\,\frac{1}{\sinh[\pi t T - i \zeta]} \nonumber \\
&=& \frac{1}{2}\delta(t)+\frac{T}{2i}\, \mathcal{P} \frac{1}{\sinh[\pi t T ]}, \label{ga2}
\end{eqnarray}
where on the second line $\zeta \to +0$ serves to regularize the integral for $t \to 0$, 
and we employed the Sokhotski-Plemelj formula with $\mathcal{P}$ denoting the principal value. 
Equation~(\ref{ga2}) shows that the real part of the bath correlator $K(t)$ relaxes very quickly (i.e. during time $t \sim \zeta \to 0$), 
while the imaginary part relaxes with a characteristic time $\hbar/(k_{\rm B}T)$; 
see \cite{martin} for further details on the microscopic realization of the Glauber transition rates. 

The quantity $\zeta^{-1}$ corresponds to the bandwidth (or the cut-off frequency) of the bath spectrum
which is normally large in the thermodynamic limit \cite{breuer}. 
In particular, it should be much larger than the energy differences of the levels of the system coupled to the bath. 
In the main text, we employ a logarithmic spectrum of the system with the largest energy difference being proportional to $a\ln [N]$.
We then get from $a \zeta \ln [N] \ll 1$ the following limitation on the number of energy level $N$:
\begin{equation}
\label{tarzan}
N\ll e^{1/(a\zeta)}. 
\end{equation}
Given a sufficiently small value of $a\zeta$, there is a large room for satisfying (\ref{tarzan}). 
This is the advantage of using the logarithmic spectrum.

\subsection{Stinespring theorem}
\label{app:s12}

Given non-trivial conditions needed to derive a dissipative Markovian dynamics 
from the weak-coupling system-bath approach, one may want to compare 
the dissipative dynamics of a system with a global unitary evolution. 
To this end, we recall the Stinespring theorem \cite{breuer,chuang}:
dissipative dynamics that operate on diagonal density matrices correspond to
a completely positive map (CPM). Any CPM on a $N$-dimensional Hilbert space 
$\mathcal{H}_{N}$ can be represented as a partial trace of a unitary operator $U$ 
acting on an $N^3$-dimensional Hilbert space $\mathcal{H}_{N^3}=\mathcal{H}_{N} \otimes \mathcal{H}_{N^2}$. 

The Stinespring theorem provides a finite-dimensional environment for unitary modeling of a CPM. 
This conforms to the standard set-up of quantum computation \cite{chuang} and contrasts with 
the system-bath approach that generally involves infinite-dimensional (due to the thermodynamic limit) bath models. 

The Stinespring theorem, however, cannot be applied to our dissipative search procedure to compare its complexity 
to that of the unitary Grover search. Indeed, if a unitary evolution $U$ is to be related to the search problem, 
the Hamiltonian $H_{\mathrm{Stin}}$ that generates $U$ in $\mathcal{H}_{N^3}$ should be of form 
$H_{\mathrm{Stin}} = H_{\mathrm{S}} +H_{\mathrm{E}} + V_{\mathrm{I}}$, where both the environment Hamiltonian $H_{\mathrm{E}}$ living in $\mathcal{H}_{N^2}$
and the interaction Hamiltonian $V_{\mathrm{I}}$ living in $\mathcal{H}_{N^3}$ should not dependent on the unknown state $\ket{w_{\ell}}$ 
and thereby the Hamiltonian $H_{\mathrm{S}}=H_0+H_1$ of the system. 
Otherwise, if the system Hamiltonian, and thereby its ground state $\ket{w_{\ell}}$, were known, there is nothing to search,
and if a unitary transformation should be designed to rotate the state vector from an initial well-defined state, 
such as, e.g., $\ket{s} = \frac{1}{\sqrt{N}}\sum_{k=1}^N \ket{w_k}$, to the (known) final state $\ket{w_{\ell}}$,
it can be done optimally with a complexity $\mathcal{O}(1)$ for any given $N$. Hence, the problem in applying $U$
is that it is not generated by a legitimate search Hamiltonian. 

Another hindrance is that $H_{\mathrm{Stin}}(t)$ is generally not time-independent \cite{burgath}.  
Moreover, we cannot implement $H_{\mathrm{Stin}}(t)$ via external sources, because it would have to
depend on the unknown state $\ket{w_\ell}$. In contrast, the system-bath approach employs a time-independent Hamiltonian. 

Hence, the Stinespring theorem does not allow one to conclude that the dissipative search described by a Markovian dynamics
should be computationally less efficient than the unitary Grover search in $\mathcal{H}_{N}$. 

\section{Formal solution of the master equation}
\label{app:s2}

We write the master equation for the vector $\ket{p}$ of populations of states $\{ \ket{w_k} \}_{k=1}^N$
using the Dirac notation,
\begin{equation}
\partial_t \ket{p} = A \ket{p} , \label{appeq:ME}
\end{equation}
where the matrix elements $A_{kl} \equiv v_{kl} - \delta_{kl} \sum_{j=1}^N v_{jk}$, with $v_{kk} =0$, 
satisfy $A_{k\not=l}\geq 0$ and $A_{kk} = - \gamma_k < 0$.
Note that, due to the detailed balance condition,  
$\tilde{A}_{kl}=A_{kl}e^{-\frac{\beta}{2}(\epsilon_l-\epsilon_k)}=\tilde{A}_{lk}$
is a symmetric matrix that has the same eigenvalues as $A_{kl}$. 
Hence, $A_{kl}$ is a diagonalizable matrix with properly defined (orthonormal) left and right eigenvectors.
The right eigenvector $\ket{R_1}$ of $A$ with eigenvalue $\alpha_1 = 0$ coincides 
with the stationary Gibbsian probability \cite{kampen}: $\sum_{l=1}^N A_{kl}e^{-\beta \epsilon_l}=0$. 
The corresponding left eigenvector $\bra{L_1}$ has all its components equal to $1$, as seen from $\sum_{k=1}^N A_{kl}=0$.
Writing the eigenresolution of $A$ as
\begin{gather}
 A = \sum_{k=1}^N \alpha_k \ket{R_k} \bra{L_k} , \quad \braket{L_k| R_l } = \delta_{kl}, \\
 A \ket{R_k} = \alpha_k \ket{R_k} , \quad \bra{L_k} A = \alpha_k \bra{L_k} , 
\end{gather} 
where $\{ \ket{R_k} \}_{k=1}^N$ and $\{ \bra{ L_k } \}_{k=1}^N$ are the right and left eigenvectors, 
we can formally solve Eq.~(\ref{appeq:ME}) via 
$e^{At} = \sum_{k=1}^N e^{\alpha_kt} \ket{R_k} \bra{L_k}$ with $\alpha_1 = 0 > \alpha_2 \geq \alpha_3, \ldots$,
leading to \cite{kampen} 
\begin{eqnarray}
\ket{ p(t) } &=& e^{At} \ket{ p(0) } 
\nonumber \\ 
& \simeq & \ket{p^{(\mathrm{st})}} + e^{-|\alpha_2| t } \braket{L_2|p(0)} \ket{R_2} + \mathcal{O} [e^{-|\alpha_3|t}] , \quad 
\end{eqnarray}
where $\ket{ p^{(\mathrm{st})}} = Z^{-1} \sum_{k=1}^N e^{-\beta\varepsilon_k} \ket{w_k}\bra{w_k}$, 
with $Z = \sum_{k=1}^N e^{-\beta\varepsilon_k}$, is the stationary state. 
We can therefore define the relaxation time as $\tau_{\mathrm{rlx}} = 1/|\alpha_2|$.

Spectral features of Markov matrices are needed in many applications and are
extensively studied; see, e.g. \cite{chen}.

\section{Glauber rates for the logarithmic spectrum}
\label{app:s3}

In the main text, we employ the Glauber rates 
\begin{equation}
v_{kl} = \frac{v}{\max (n_k, n_l)} \, \left(1+e^{-\beta \varepsilon_l+\beta \varepsilon_k} \right)^{-1}, \label{appeq:Grauberv}
\end{equation}
for the transitions between the states with energies $\varepsilon_k$ and $\varepsilon_l$, and here 
we present the corresponding explicit expressions for the logarithmic spectrum $\eta_k = a \ln(k)$ 
of the auxiliary Hamiltonian $H_1$. Consider first the case of the ground state of $H_0$ at $\ell=1$
with energy $\epsilon = -b \ln(N)$. The energy levels of $H = H_0 + H_1$ are 
$(\varepsilon_1, \varepsilon_2, \dots ,\varepsilon_N ) =(-b \ln N, a \ln 2,..., a \ln N)$,
and the transition rates (\ref{appeq:Grauberv}) are 
\begin{subequations}
\begin{eqnarray}
v_{1l} &=& \frac{1}{l [1+l^{-a\beta }N^{-b\beta}]}, \quad l>1 , \\
v_{l1} &=& \frac{1}{l [1+l^{a\beta}N^{b\beta}]}, \quad l>1 , \\
\frac{v_{1l}}{v_{l1}}&=&l^{a\beta}N^{b\beta}, \quad l>1 , \\
v_{k<l} &=& \frac{1}{l [1+(k/l)^{a\beta}]}, \quad l>1, \, k>1 ,\\ 
v_{l>k} &=& \frac{1}{l [1+(l/k)^{a\beta }]}, \quad l>1, \, k>1, \\
\frac{v_{k<l}}{v_{l>k}}&=& \left( \frac{k}{l} \right)^{-a\beta}, \quad l>1, \, k>1 
\end{eqnarray}
\end{subequations}
Note that for high temperatures $a\beta < 1$ the transition rate $v_{lk}$ from 
a state with lower energy $\varepsilon_k$ to a state with higher energy $\varepsilon_l$ is smaller, 
but comparable with the reverse transition rate $v_{kl}$, which leads to longer
relaxation times $\tau_{\mathrm{rlx}}$ as is also confirmed numerically. 
The total transition rate from state $\ket{w_l}$ is then 
\begin{subequations}
\begin{eqnarray}
\gamma_1 &=& \sum_{s=1}^N v_{s1} = \sum_{s=2}^{N}s^{-1}\left[1+N^{b\beta} s^{a \beta}\right]^{-1}, \\
\gamma_l &=& \sum_{s=1}^N v_{sl} = \frac{1}{l}\left[1+  N^{-b\beta}l^{-a \beta}\right]^{-1}
+\frac{1}{l}\sum_{s=2}^{l-1} \left[ 1 +(s/l)^{a \beta}\right]^{-1} 
\nonumber \\ & &
+ \sum_{s=l+1}^{N}s^{-1}\left[1+  (s/l)^{a \beta}\right]^{-1}, \quad l>1, 
\end{eqnarray}
\end{subequations}
and $\gamma_l \lesssim 1$ holds for any $a \beta > 0$, as required. 

The same conclusions hold for other values of $\ell$, e.g., for $\ell=N$ the energy levels are 
$( \varepsilon_1, \varepsilon_2, \dots ,\varepsilon_N ) = \big( (a-b) \ln N, \, 0, \, a\ln 2, \ldots, a\ln (N-1) \big)$
and the transition rates are 
\begin{subequations}
\begin{eqnarray}
v_{1l} &=& \frac{1}{l [1+(l-1)^{-a\beta}N^{- (b-a) \beta}]}, \quad l>1, \\
v_{l1} &=& \frac{1}{l [1+ (l-1)^{ a \beta}N^{ (b-a) \beta}]}, \quad l>1, \\
\frac{v_{1l}}{v_{l1}} &=& (l-1)^{a\beta} \, N^{(b-a)\beta}, \quad l>1, \\
v_{k<l} &=& \frac{1}{l [1+ ( (k-1)/(l-1) )^{\beta a}]}, \quad l>1,\, k>1, \qquad \\
v_{l>k} &=& \frac{1}{l [1+ ( (l-1)/(k-1) )^{\beta a}]}, \quad l>1, \, k>1, \\
\frac{v_{k<l}}{v_{l>k}} &=& \left(\frac{k-1}{l-1}\right)^{-\beta a}, \quad l>1, \, k>1 ,
\end{eqnarray}
\end{subequations}
and for $\varepsilon_k < \varepsilon_l$ we again have $v_{lk} < v_{kl}$, provided $b>a$ which is always assumed. 
The total transition rate from any state $\ket{w_l}$ is  
\begin{subequations}
\begin{eqnarray}
\gamma_1 &=&  \sum_{s=1}^N v_{s1} = \sum_{s=2}^{N}s^{-1}\left[1+N^{(b-a)\beta}(s-1)^{a \beta}\right]^{-1} , \qquad \\
\gamma_l &=& \sum_{s=1}^N v_{sl}=\frac{1}{l}\left[1+  N^{(a-b)\beta}(l-1)^{-\beta a}\right]^{-1}  
\nonumber \\ & & 
\frac{1}{l}\sum_{s=2}^{l-1}\left[1+  \left( \frac{s-1}{l-1}\right)^{a \beta}\right]^{-1}
\nonumber \\ & &
+\sum_{s=l+1}^{N}\frac{1}{s}\left[1+  \left( \frac{s-1}{l-1}\right)^{a \beta}\right]^{-1} .
\end{eqnarray}
\end{subequations}

\section{Formulas}

\begin{eqnarray}
\geq \frac{\pi\hbar}{2}\left[\langle\psi| H_1^2|\psi\rangle-\langle\psi| H_1|\psi\rangle^2   \right]^{-\frac{1}{2}},\\
\frac{\pi\hbar\sqrt{N}}{2|\varepsilon|}
\end{eqnarray}

\begin{eqnarray}
p_k=\langle w_k |\rho|w_k\rangle\propto e^{-\varepsilon_k/T}
\end{eqnarray}

\begin{eqnarray}
v_{kl}=0\quad |k-l|\geq 2
\end{eqnarray}

\begin{equation}
\sum_{l=1}^{N} v_{lk} \lesssim v=O(1) \quad \forall \; k, 
\end{equation}

\begin{equation}
v_{lk}>0
\end{equation}

\begin{equation}
v_{kl} = \frac{v}{\max (n_k, n_l)} \, \left(1+e^{-(\varepsilon_l- \varepsilon_k)/T} \right)^{-1}, 
\end{equation}

\begin{equation}
n_k=\{ \# {\rm levels}\leq \varepsilon_k   \}
\end{equation}

\begin{equation}
v_{kl}~e^{-\varepsilon_l/T}=v_{lk}~e^{- \varepsilon_k/T}
\end{equation}

\begin{equation}
\varepsilon_1=-b\ln N \qquad \varepsilon_{k\geq 2}=a\ln k
\end{equation}

\begin{equation}
b>a>T  \qquad \tau_{\rm relaxation}\simeq O(\ln [N])
\end{equation}



\begin{thebibliography}{99}
\bibitem{comment_unstructured}
When searching an unordered set of $N$ elements, in the best (worst) case 
the desired element will be found in the first ($N$th) step and all cases are equally probable,
leading on average to $(N+1)/2$ steps. 
But if some order is already present in the set, the classical search can proceed much 
faster than in $\mathcal{O}(N)$ steps. Indeed, assume that the elements are indexed by numbers
$i_1,i_2,\ldots,i_N$ ordered as $i_1 \leq i_2 \leq \ldots \leq i_N$. 
These numbers are not known, apart from $i_\ell$ for the desired element, but we do not know its position $\ell$. 
Then $i_\ell$ can be found in $\mathcal{O}(\log_2 N)$ steps by dividing the set into two halves, 
checking an element at the intersection of the two halves and selecting 
one of the halves for further interrogation, and so on.
\bibitem{comment_sstate}
For $n$ qubits, $N=2^n$, state $\ket{s}$ can be prepared locally 
as a tensor product state $\ket{s} = \left( \frac{\ket{0} + \ket{1}}{\sqrt{2}} \right)^{\otimes n}$.
\end{thebibliography}
\end{document}